\documentclass[english]{revtex4-1}
\usepackage[T1]{fontenc}
\usepackage[latin9]{inputenc}
\usepackage{geometry}
\usepackage{textcomp}
\usepackage{amsmath}
\usepackage{amssymb}
\usepackage{stackrel}
\usepackage{esint}

\makeatletter

\usepackage{babel}

\makeatother

\usepackage{babel}
\begin{document}

\title{Analysis of the Jun Ishiwara's \textquotedbl{}The universal meaning
of the quantum of action\textquotedbl{}}

\author{Karla Pelogia}

\affiliation{Philosophisch - Historische Fakultät, Universität Stuttgart Keplerstr. 17,
70174 Stuttgart, Deutschland}
\email{karlapelogia@gmail.com}

\author{Carlos Alexandre Brasil}

\affiliation{São Carlos Institute of Physics (IFSC), University of São Paulo (USP),
PO Box 369, 13560-970 São Carlos, SP, Brazil}
\email{carlosbrasil.physics@gmail.com}

\begin{abstract}
Here we present an analysis of the paper ``Universelle Bedeutung
des Wirkungsquantums\textquotedblright{} (The universal meaning of
the quantum of action), published by Jun Ishiwara in German in the
``Proceedings of Tokyo Mathematico-Physical Society 8 (1915) 106-116\textquotedblright .
In his work, Ishiwara, established in the Sendai University, Japan,
proposed - simultaneously with Arnold Sommerfeld, William Wilson and
Niels Bohr in Europe - the phase-space-integral quantization, a rule
that would be incorporated into the old-quantum-theory formalism.
The discussions and analysis render this paper fully accessible to
undergraduate students of physics with elementary knowledge of quantum
mechanics. 
\end{abstract}
\maketitle

\section{Introduction}

No theory defies our common sense as much as quantum mechanics (QM).
Richard Feynman (1918-1988) said explicitly on his lecture \textquotedbl{}Probability
and Uncertainty: The Quantum Mechanical View of Nature\textquotedbl{}
\cite{Feynman} that nobody understands the theory, Mário Schenberg
(1914-1990) said that QM is the most important scientific revolution
of the history of humanity \cite{Schenberg} and the debates between
Albert Einstein (1879-1955) and Niels Bohr (1885-1962) about the interpretation
of QM are notorious (see the first chapters of \cite{Zurek}). The
QM reached its standard form with the modern wave theory based on
the Schrödinger (presented on his set of papers \textquotedbl{}Quantization
as a problem of proper values I-IV\textquotedbl{} \cite{Ludwig,Schrodinger})
and Dirac \cite{Dirac1} equations, with the probabilistic interpretation
for the wave function by Max Born (1882-1970) (presented on his papers
\textquotedbl{}On the quantum mechanics of collisions\textquotedbl{}
\cite{Zurek} and \textquotedbl{}Quantum mechanics of collision processes\textquotedbl{}
\cite{Ludwig}), and Paul Dirac (1902-1984) \cite{Dirac2}. However,
the path since the establishment of the concept of energy quantum
by Max Planck (1858-1947) (translations from german of his original
papers can be found on \cite{Haar,Planck1,Planck2} and analysis about
them on \cite{Feldens,Studart} ) \textendash{} in a process described
by himself as ``\emph{an act of desperation }{[}done because{]}\emph{
}a theoretical explanation\emph{ }{[}to the black-body-radiation spectrum{]}
had to be supplied\emph{ at all cost, whatever the price}\textquotedblright{}
\cite{Jammer} \textendash{} until the wave equations lasted about
3 decades \cite{Gamow,Jammer}.

During that time, the physics of the atomic phenomena evolved into
a theory now known as the \emph{old quantum theory} (OQT) \cite{Bucher,Jammer,Tomonaga},
which was based on a heuristic approach to atomic phenomena according
to the following rules \cite{Jammer}: 
\begin{itemize}
\item i) the use of classical mechanics to determine the possible motions
of the system; 
\item ii) the imposition of certain quantum conditions to select the actual
or allowed motions; 
\item iii) the treatment of radiative processes as transitions between allowed
motions, subject to Bohr's frequency formula. 
\end{itemize}
From the third rule, we can see the important role played by Bohr
\cite{Kragh,Parente,Svidzinsky} with his atomic model\cite{Bohr1,Bohr2,Bohr3,Calouste},
who spurred one of the greatest achievements in physics during the
first decades of the twentieth century: the theoretical derivation
of Balmer's formula \cite{Banet1,Banet2}. The Bohr model was based
on the following postulates (as originally published in \cite{Bohr1}): 
\begin{enumerate}
\item \textquotedbl{}That the dynamical equilibrium of the systems in the
stationary states can be discussed by help of the ordinary mechanics,
while the passing of the systems between different stationary states
cannot be treated on that basis; 
\item That the latter process is followed by the emission of a \emph{homogeneous}
radiation, for which the relation between the frequency and the amount
of energy emitted is the one given by Planck's theory.\textquotedbl{} 
\end{enumerate}
We can see that the third rule of the OQT was based on Bohr's second
postulate. At last, in the calculations, an adequate version of the
\emph{correspondence principle}, formulated in distinct forms by Bohr,
Planck and Werner Heisenberg (1901-1976)\cite{Darrigol,Jammer,Liboff,Makowski},
which states that when we take the limits of some basic parameters
of QM we must find the classical results again \cite{Bhattacharyya,Hassoun,Jammer,Liboff,Makowski}.
The pioneer example of the application of the correspondence principle
was made by Bohr \cite{Parente} on Sec. 3 (\textquotedbl{}General
considerations continued\textquotedbl{}) of Ref. \cite{Bohr1}, where
he uses it to obtain the correct dependence between the electronic
frequencies of emission/absorption and translation around the nucleus.

An improvement of the OQT would be attained through phase-space analysis,
something that is not surprising, since Planck's constant $h$ (see
translations of original Planck's papers \textquotedbl{}On an improvement
of Wien's equation for the spectrum\textquotedbl{} and \textquotedbl{}On
the theory of the energy distribution law of the normal spectrum\textquotedbl{},
with historical analysis, on \cite{Haar,Planck1,Planck2}) has the
dimension of an action (and angular \emph{momentum}) \cite{Bucher,Eckert,Feldens,Studart}.
New conditions for quantization then emerged, proposed by William
Wilson (1875-1965), Jun Ishiwara (1881-1947), Arnold Sommerfeld (1868-1951)
and Bohr. Bohr's (see the first part of his paper \textquotedbl{}On
the quantum theory of line-spectra\textquotedbl{} on \cite{Waerden})
and Sommerfeld's \cite{Sommerfeld3,Sommerfeld4,Sommerfeld1,Sommerfeld2}
contributions have been subject of detailed scrutiny - for the recent
literature see \cite{Aaserud,Duncan,Eckert,Kragh2,Seth} - while Wilson's
\cite{Wilson,Wilson2} and Ishiwara's \cite{Ishiwara,Ishiwara3} papers
still await historical analysis.

The original papers of Wilson, Bohr and Sommerfeld are accessible
in English. In the case of Sommerfeld, see \cite{Sommerfeld1,Sommerfeld2};
Sommerfeld's book \emph{Atombau und Spektrallinien} - then regarded
as \emph{the Bible of quantum spectroscopy} (see \cite{Duncan,Seth}
and the Eckert's contribution on \cite{Aaserud}) - was translated
already in 1923 into English \cite{Sommerfeld5}. The same has not
happened to the work of Ishiwara, \textquotedbl{}Universelle Bedeutung
des Wirkungsquantums\textquotedbl{}, published in German in the \emph{Proceedings
of Tokyo Mathematico-Physical Society} \cite{Ishiwara}. This oversight
is being rectified by the authors, presenting here for the first time
a translation into English \textquotedbl{}The universal meaning of
the quantum of action\textquotedbl{} \cite{Ishiwara3}.

Here, we present an analysis of Ishiwara's paper, with a brief biographic
section, historical precedents on QM and a comparison among the approaches
of Ishiwara, Wilson, Bohr and Sommerfeld. We intend this paper to
be useful not only to researchers of history of science, but also
to undergraduate students of physics and students of other exact sciences
interested to understand the development of the modern physical ideas,
who can learn more about the phase-space analysis, the treatment of
elliptic motion on OQT and the development of quantum ideas, e.g.
the photoelectric effect and atomic models. It requires only a minimal
knowledge of QM or modern physics, at the level of Ref. \cite{Eisberg},
and the present paper can be complemented with \cite{Castro}. We
were careful to choose the references by its original value and didactic
character, besides those important for our principal aim of historical
analysis.

The present paper is organized as follows: in section 2 we present
a short profile of Jun Ishiwara; in section 3 we make a brief discussion
about the precedents on OQT to the work of Ishiwara and an analysis
of each section of his paper; in section 4 we present our conclusions.

\section{A brief biography of Jun Ishiwara}

Jun Ishiwara (1881-1947) \textendash{} his name is sometimes translated
as Ishihara \textendash{} was educated at the Tokyo Imperial University,
graduating in theoretical physics in 1906. He was, together with Yoshio
Nishina (1890-1951), one of Hantaro Nagaoka's students (1865-1950)
\cite{Frederic} - Nagaoka was the author of the \emph{Saturnian atomic
model} \cite{Nagaoka}. In 1908, Ishiwara became a teacher at the
Army Gunnery Engineering School \cite{Sigeko} and in 1911 he was
appointed assistant professor at the Graduate School of Science and
Faculty of Science, Tohoku University, Sendai \cite{Mehra}. From
1912 to 1914, Ishiwara studied in Europe at the University of Munich,
the E. T. H. (Eidgenössische Technische Hochschule) in Zurich and
at the University of Leyden, studying with Sommerfeld and Einstein
\cite{Einstein1,Mehra}. After returning to Japan, he was named for
the chair of Full Professor of Physics at Tohoku University, in Sandai
and, in 1919, he received, for his works on relativity and QM, the
Gakushiin award by the Imperial Academy of Sciences \cite{Einstein1,Frederic,Sigeko}.

Ishiwara dedicated most of his time to Einstein's theory of relativity,
and he was a scholar and debater of the theory, gaining reputation
both in Japan and China, where he was known as ``\emph{the only expert
of relativity studies in Japan} (\dots )\emph{ }{[}who{]}\emph{ }had
a unique understanding of the theory of relativity\textquotedblright \cite{Hu}.
The first paper by Ishiwara on the theory was published in 1909 and
was the first in Japan on this topic \cite{Sigeko}.

When, in late 1922, the publishing company Kaizosha invited Einstein
to visit Japan \cite{Low}, it was Ishiwara who hosted him \cite{Hu,Low}.
Einstein arrived in Japan on November 17 and stayed for six weeks\cite{Low,Pais}.

With an invitation by Kitaro Nishida (1870-1945), professor of Philosophy
at Kyoto University, Einstein gave a talk entitled ``How I created
the theory of relativity\textquotedblright , when he clarified several
aspects of his creative process and the route to his main works. Einstein
made no written notes for the talk, however, and the only record of
Einstein's words are the careful notes taken by Ishiwara, who also
provided a running translation from German to Japanese\cite{Einstein1,Pais}.
In fact, it was in his stay in Japan that Einstein received the news
of his Nobel Prize \cite{Einstein1,Einstein2}. Kaizosha's collection
of Einstein's complete works was the first in the world, selling about
4000 copies, and was edited by Ishiwara \cite{Hu,Low}.

There are several letters between both physicists in Einstein's archive
\cite{Einstein3}. For example, on January 12th 1923, soon after Einstein's
visit, Ishiwara wrote to Einstein beginning with: 
\begin{quote}
``Esteemed Professor, Your stay in our country was a special pleasure
for me that I shall always cherish as such a fine memory. You are
probably still happily continuing your journey!\textquotedbl{} 
\end{quote}
followed by a discussion on tensorial analysis \cite{Ishiwara2}.

On QM the first contribution of Ishiwara was published in the inaugural
volume of\emph{ The science reports of the Tohoku University}\cite{Sigeko}.
He was studying the electron theory on metals, a research - on Sigeko's
words \cite{Sigeko} - \textquotedbl{}important in that it paralleled
Bohr's degree thesis of 1911 on the electron theory of metals and
his subsequent well-known quantum theory of atomic structure\textquotedbl{}.
As noted by Abiko \cite{Abiko}: 
\begin{quote}
\textquotedbl{}In 1911, {[}Ishiwara{]} submitted (though published
in 1912) his own original paper in German 'Contribution to light-quantum
theory', in which he deduced Planck's radiation formula from the viewpoint
of radiation as a collection of light-quanta, in a similar way to
Bose's statistics of 1924 (...). Also in that paper of 1911, similarly
to de Broglie did by way of introducing the phase waves in 1923, Ishiwara
tried to explain wave-like behaviour of radiation as a collection
of light-quanta, by associating minute electric and magnetic vectors
to each light quantum.\textquotedbl{} 
\end{quote}
We see here that Ishiwara anticipated by almost 15 years fundamental
ideas of Satyendra Nath Bose (1894-1974) and Louis de Broglie (1892-1987)\cite{Abiko}.
In another work in the same year, Ishiwara supported the light-quantum
hypothesis about the constitution of x-rays and $\gamma$-rays in
an pioneering attitude, against the opinion of the Braggs - William
Henry (1862-1942) and William Lawrence (1890-1971) - for who these
rays were constituted by neutral particles. To emphasizing the pioneerism
of Ishiwara, it is important to say that Einstein's light quantum
theory, at the time, was supported only by a minority of physics \cite{Abiko,Sigeko}.
In his 1915 paper analyzed here, he suggested the application of the
phase-space-integral quantization to the hydrogen atom, supposing
elliptical orbits \cite{Ishiwara,Ishiwara3} - that was the first
published reference to Bohr's theory by a non-western physicist \cite{Kragh2}.

By 1918, 4 years before Einstein's visit to Japan, Ishiwara unfortunately
diminished his research activity \cite{Sigeko,Tabata}. He took leave
from the university in 1921 and formally retired in 1923, working
then mainly in divulgation and journalism of science \cite{Abiko}
- indeed, he was pioneer in those fields in Japan\cite{Tabata} -
and dedicated himself to \emph{tanka} (a type of short Japanese poetry\cite{Frederic}).
He continued to influence many young physicists with his literary
and scientific divulgation activities \cite{Sigeko}. At last, two
treatises of Ishiwara, \textquotedbl{}The fundamental problems of
physics\textquotedbl{}, were published in his retirement time - the
first volume on 1923 and the second on 1926 - including sections about
quantum theory and atomic constitution. These books were references
among the specialists of physics\cite{Abiko}.

\section{Analysis}

\subsection{Precedents}

The analysis of phase-space has been conducted over many years, before
the Ishiwara work. Three forms of quantization for the phase-space
were proposed at the first Solvay Congress \cite{Jammer,Straumann}.
It is important, here, to emphasize that the 'phase-space' expression
was coined by Paul Ehrenfest (1880-1933) in a review paper of the
work of Ludwig Boltzmann (1844-1906), published on 1911 \cite{Nolte}.
Because it is a relatively recent expression at the time, we should
not be surprised that Ishiwara did not used it on his paper, writing
\textquotedbl{}state space\textquotedbl{} instead. However, in our
discussions here, we will use the now familiar and consolidated expression,
independently of its use (or not) by the followed mentioned authors.

Planck \cite{Eckert,Jammer,Straumann} suggested the rule defined
by 
\begin{equation}
\int\int dp\,dq=h\label{Planck}
\end{equation}
where, and this elicited a discussion with his colleagues, including
Sommerfeld and Einstein. Following the Planck presentation, Henri
Poincaré (1854-1912) \cite{Sigeko} suggested the expansion of Planck's
rule, for a system with $n$ degrees of freedom, to

\begin{equation}
\int\int\left(dp_{1}\,dq_{1}+dp_{2}\,dq_{2}+...+dp_{n}\,dq_{n}\right)=h\label{Poincare}
\end{equation}

By his time, Sommerfeld presented in his works in the field that,
for every molecular process, the quantity of action

\begin{equation}
\int_{0}^{\tau}L\,dt=\frac{h}{2\pi}\label{Sommerfeld}
\end{equation}
where $\tau$ is the duration of the process and $L$ is the Lagrangian.
This is the same form presented on the Karlsruhe meeting \cite{Sommerfeld6},
originating discussions with other participants on both events \cite{Jammer,Straumann}.
The precedents for the Sommerfeld formula of 1911 were his works on
a theory about $\gamma$-rays and $\beta$-decay, where both emissions
were mutually related - it is convenient to emphasize that, at the
time, there was no nuclear theory yet since Rutherford published the
results of his pioneering experiments of scattering during 1911-1914
\cite{Herman,Jammer}. Prior to (\ref{Sommerfeld}), after an exchange
of letters with Planck, Sommerfeld wrote the expression

\begin{equation}
\int_{0}^{\tau}H\,dt=\frac{h}{2\pi}
\end{equation}
where $H$ was the dynamical potential and for \textquotedbl{}the
most part we shall view $H$ as the mere abbreviation for $T-V$\textquotedbl{},
where $T$ and $V$ are the kinetic and potential energy, respectively
\cite{Herman}. Then, on 1913, were published the classic set of papers
\cite{Bohr1,Bohr2,Bohr3} where Bohr presents his atomic model.

These are the precedents for the final form of the phase-space quantization
rule, 
\begin{equation}
\int q_{i}\,dp_{i}=n_{i}\,h\label{aux2}
\end{equation}
where $q$ indicates the position and the $p$ indicates the \emph{momentum}.

With these pieces, it is not difficult to have a view of the route
to the works of Wilson, Ishiwara, Sommerfeld and the subsequent work
of Bohr.

The first author to present the definitive rule was Wilson \cite{Mehra2,Mehra}
in a paper for the \emph{Philosophical Magazine} from 1915 \cite{Wilson},
communicated by John William Nicholson (1881-1955) - the priority
of discovery was, indeed, recognized by Sommerfeld \cite{Mehra2}.
Wilson wanted to establish the \textquotedbl{}possibility of deducing
the results of Planck and Bohr from a single form of quantum theory\textquotedbl{}\cite{Wilson}
and he made his theory with the hypothesis of steady states of a system,
that can be analyzed by Hamiltonian dynamics behind (\ref{aux2}),
and postulating the existence of discontinuous process between these
steady states with absorption or emission of energy. With this he
obtained, in his first paper \cite{Wilson}, Bohr's formula for the
kinetic energy of the electron and the Planck distribution law. In
an other paper \cite{Wilson2}, also in \emph{Philosophical Magazine},
Wilson stated more clearly the assumptions of his theory and developed
it for the emission of spectral lines, citing both the rule (\ref{Planck})
and the work of Ishiwara, published between Wilson's two works.

By other side, Sommerfeld began to work in Bohr's model, as described
by Eckert \cite{Eckert}, motivated by his quest for a theory of the
Zeeman and Stark effects \cite{Eckert,Jammer}. Indeed, \textquotedbl{}Sommerfeld's
response to Bohr's atomic model was the earliest reaction from outside
Rutherford's circle, where Bohr had spent some time as a 'postdoc',
and it revealed a vivid interest in the theory of the {[}at the time{]}
unknown author\textquotedbl{} \cite{Eckert}. Sommerfeld was impressed
with the expression founded by Bohr for the Rydberg constant in terms
of fundamental parameters (in Sommerfeld's own words, 'calculating
this constant is undoubtedly a great feat' \cite{Eckert}), and was
intrigued with the possibilities concerning the Zeeman and Stark effects
\cite{Eckert,Jammer}.

During the First World War (from which Sommerfeld was dispensed),
Sommerfeld corresponded with Karl Schwarzschild (1873-1916) - who
died from a skin disease that he brought home from the front (see
the bio profile on \cite{Voigt}) - Friedrich Paschen (1865-1947),
Wilhelm Lenz (1888-1957) and Wilhelm Wien (1864-1928) and soon he
become convinced of the consistency of Bohr's model. To explain the
decomposition of spectral lines in electric fields published by Stark
in 1914, Sommerfeld proposed the existence of elliptic orbits for
the electrons, accompanying the circular Bohrian orbits \cite{Eckert,Jammer}.
New developments on hydrogen and X-ray spectra culminated on his treatises
presented to the Bavarian Academy of Science on December, 1915 \cite{Sommerfeld1}
and January, 1916 \cite{Sommerfeld2}. The treatment of two-dimensional
motion of the electron in its orbital plane indicated how systems
with several degrees of freedom could be approached, at first with
the rule (\ref{Planck}) of Planck in 1911 \cite{Eckert,Sommerfeld1,Sommerfeld2};
soon after, having recognized that the quantization of the angular
\emph{momentum} could be expressed by requiring that

\begin{equation}
\oint p_{\varphi}d\varphi=n_{\varphi}h
\end{equation}
where $p_{\varphi}$ is the angular (not the \emph{linear}) \emph{momentum}
associated to the azimuthal angle $\varphi$, with the integration
extended over the period, Sommerfeld postulated the \textquotedbl{}final\textquotedbl{}
rule (\ref{aux2}). His ideas were detailed in a treatise for the
\emph{Annalen der Physik}, \textquotedbl{}On quantum theory of spectral
lines\textquotedbl{} \cite{Sommerfeld3,Sommerfeld4} and, finally,
presented in his classical book \cite{Sommerfeld5}. Sommerfeld refers
in his papers \cite{Sommerfeld3,Sommerfeld4,Sommerfeld1,Sommerfeld2}
to the Planck works, but did not refer to the earlier publications
of Ishiwara and Wilson. This omission was corrected in his book \cite{Sommerfeld5}.
His work received major notoriety, because he made the theory more
robust and detailed \cite{Jungnickel}, generalized the quantum law
of phase integrals to apply to systems of any number of degrees of
freedom, revealing the close connection of the phase integrals with
the Hamilton-Jacobi theory and applying to Bohr's atom (a deep discussion
about his calculations can be found on \cite{Castro}), considering
elliptical orbits that allowed to deduce not just Balmer's rule but
the formula for the fine structure of hydrogen spectra. It was considered
the highest point of OQT \cite{Bucher,Jammer}, referred to by Planck
as \textquotedbl{}an achievement fully comparable with that of the
famous discovery of the planet Neptune whose existence and orbit was
calculated by Leverrier before the human eye had seen it\textquotedbl{}
\cite{Planck4}.

At last, is convenient to mention that the meaning of the Planck's
constant has changed to Sommerfeld between 1911 and his final communications.
When researching the $\gamma$ and $x$-rays (at the time, named \textquotedbl{}Röntgen
rays\textquotedbl{} \cite{Sommerfeld1}), he postulated in 1911 -
he remebered it in his 1915 presentation \cite{Sommerfeld1} - the
relation $Energy\,\times\,Time=h$ as general, but on 1915, he restricted
the hypothesis \textquotedbl{}essentially to periodic motions, and
envisages an integral multiple of $h$, furthermore equality is replaced
by a proportionality depending on the force law\textquotedbl{} \cite{Sommerfeld1}.
It was 1911 Sommerfeld's hypothesis that Ishiwara used in the analysis
of his paper\cite{Ishiwara,Ishiwara3}.

The last work considered here is the Bohr one, the communication ``On
the quantum theory of line-spectra\textquotedblright{} published in
English in 1918 in the \emph{Mémoires de l'académie royale des sciences
et des lettres de Danemark} (the first part can be found on the reference
\cite{Waerden}). As the title itself indicates, the main concern
of Bohr is the spectral analysis. Emphasizing the limitations of classical
electrodynamics to atomic systems and remembering his own model, he
approaches systems with one degree of freedom by Hamilton dynamics.
For $s$ degrees of freedom, Bohr forms the expression

\begin{equation}
I=\int_{0}^{\sigma}\stackrel[k=1]{s}{\sum}p_{k}\dot{q}_{k}dt
\end{equation}
where $\sigma$ is the period of the motion, $q_{k}$ the coordinates
and $p_{k}$ the momentum. Analyzing small variations - \textquotedbl{}$I$
will be invariant for any finite transformation of the system which
is sufficiently slowly performed, provided the motion at any moment
during the process is periodic and the effect of the variation is
calculated on ordinary mechanics\textquotedbl{} - and making use of
Planck's theory for harmonic oscillator, Bohr arrives at

\begin{equation}
I=\int_{0}^{\sigma}p\,dq=n\,h
\end{equation}
His concern, then, is to generalize the theory, with an expression
equivalent to (\ref{aux2}), and to apply it for other periodic systems,
like a particle moving under the influence of the attractions from
two fixed centers or a particle moving in a plane executing harmonic
vibrations in two perpendicular directions with distinct frequencies.
Bohr recognizes the works of Sommerfeld and Wilson, but did not mention
Ishiwara.

Summarizing the Wilson, Sommerfeld and Bohr approaches, all of them
are based on the (\ref{aux2}) formula, the problems treated were
the hydrogen spectra with both Balmer's and the fine structure formula,
and the Planck distribution. They were all published in the period
between 1915 and 1918, but were based on discussions and works from
the beginning of the decade. After this little historical review,
the work of Ishiwara will be discussed.

\subsection{Ishiwara's \textquotedbl{}The universal meaning of the quantum of
action\textquotedbl{}}

Ishiwara's paper \cite{Ishiwara,Ishiwara3} is divided in four sections:
a short introduction, followed by the proposal of his phase-space
quantization rule, the analysis of the Bohr atomic model (perhaps
the principal part of the paper), concluding with the photoelectric
effect. The latter application was unique with the considered authors.

In the beginning, Ishiwara is concerned about the different meanings
of $h$, based, fundamentally, on the fact of the Planck\textasciiacute s
constant being identified with scalar (basic dimension on phase-space
or a parameter related to fundamental amount of energy and difference
of energy on atomic levels) and vectorial (angular momentum) quantities.
The title of the paper itself refers to the quest for an ultimate
meaning of $h$. There are two interpretations considered by Ishiwara
as opposites: the Bohr one, where $h$ is closely related to the quantized
value of angular momentum and that of Sommerfeld, where $h$ is related
to the fundamental volume of the phase-space. Moreover, he states
three questions at the beginning, answering them along the paper: 
\begin{enumerate}
\item Are those different interpretations of $h$ identical to each other?\emph{
No, because the angular momentum is not quantized a priori - its quantization
is a consequence of (\ref{eq1}) - and his final value, founded on
section B - The Bohr model of the atom, } 
\begin{equation}
f=n\frac{h}{\pi}\sqrt{1-\varepsilon^{2}}\label{eq12}
\end{equation}
\emph{- equation ($12$) on Ishiwara's paper \cite{Ishiwara,Ishiwara3}\textendash{}
for general/elliptical orbits, is distinct from the value found by
Bohr } 
\begin{equation}
f=n\frac{h}{\pi}\label{eq12linha}
\end{equation}
-\emph{ equation ($12^{\prime}$) on Ishiwara's paper \cite{Ishiwara,Ishiwara3}.} 
\item Which viewpoint of the universal meaning of $h$ is the right one?\emph{
The position of Sommerfeld on 1911, where $h$ is the fundamental
length on phase space, related to the general postulate $Energy\,\times\,Time=h$
\cite{Sommerfeld1}.} 
\item Should all phenomena be explained by one and the same basic assumption?
\emph{Yes, and the basic assumption is (\ref{eq1}).} 
\end{enumerate}

\subsubsection{The basic assumption}

The version proposed by Ishiwara to the phase-space quantization rule
is

\begin{equation}
h=\frac{1}{j}\underset{i=1}{\overset{j}{\sum}}\int q_{i}\,dp_{i}\label{eq1}
\end{equation}
That is the first equation of his paper - equation ($1$) in Ishiwara's
paper \cite{Ishiwara,Ishiwara3}\textendash{} and we can observe that
it resembles to Poincaré's version, (\ref{Poincare}), by the presence
of a sum. Unfortunately, Ishiwara does not mention the calculations
that conducted him to (\ref{eq1}) and, though there exists the possibility
that he had read the proceedings of the Solvay Congress, there is
nothing to confirm it \cite{Sigeko}.

Ishiwara cites the works of Sackur and Tetrode. In addition to the
works cited in the paper, it is also important to mention that Sackur
tried to construct a theory of gases by employing the concept of finite
cells in phase space, which size determined by Planck's constant.
He and Tetrode advanced - in parallel with Planck and Sommerfeld -
to the first attempts to construct a theory of degenerate gases, applying\emph{
chemical constant method} - a constant which occurred in the expression
for the absolute entropy of a gaseous substance, that was used by
them to the quantization of translational motion of atomic particles
on monoatomic gases and to analyze the deviation of the classical
equation at low temperatures - then anticipating the ideas of Bose
and Einstein \cite{Mehra2,Mehra}.

\subsubsection{The Bohr model of the atom}

Supposing an electron submitted to a central nuclear force, with stationary
motion, elliptical orbits and with no consideration of other electrons
of the atom, Ishiwara calculates, with his rule (\ref{eq1}), the
frequency for the emitted radiation,

\begin{equation}
\nu_{1}=\frac{2\pi^{2}m_{0}e^{4}}{h^{3}}\left(\frac{1}{n_{1}^{2}}-\frac{1}{n_{2}^{2}}\right)\label{eq21}
\end{equation}
-\emph{ }equation ($21$) on Ishiwara's paper \cite{Ishiwara,Ishiwara3}\emph{
- }\textquotedbl{}which is just what is found in Bohr's theory\textquotedbl{}\emph{.}
Curiously, Ishiwara made the hypothesis \textquotedbl{}that the central
charge of the hydrogen atom consists of two electrical elementary
ones\textquotedbl{}. Bohr later pointed out that Ishiwara's theory
was inconsistent with the usual one-electron model of neutral hydrogen
\cite{Kragh2}. This hypothesis was needed because of the form of
Ishiwara's equation (\ref{eq1}) and it was not necessary in the formulation
of Wilson, Sommerfeld and Bohr. It is appropriate, however, to question
the sources of this odd hypothesis. The most likely answer is the
work of Nicholson (who communicated the papers of Wilson)\cite{Kragh2}.

Nicholson was a student of line spectra of celestial bodies and supposed
that the matter in stars could be more elementary structures than
the known chemical atoms found on Earth. Following Thomson's model,
he proposed an atomic model based on classical mechanics and electrodynamics,
imagining primary atoms consisting of small spheres of negative electricity,
rotating around a smaller spherical positive nucleus. Some of these
structures were \textquotedbl{}coronium\textquotedbl{}, the most simple,
contained two electrons; the next, \textquotedbl{}hydrogen\textquotedbl{},
three electrons, followed by more complexes \textquotedbl{}nebulium\textquotedbl{}
and \textquotedbl{}protofluorine\textquotedbl{}. With the primary
atoms, he constructed the chemical elements, and the hydrogen \emph{element}
consisted of two atoms of \emph{primary} hydrogen \cite{Kragh2,Mehra2}.
Nicholson made criticisms about Bohr's non-classical model and theory,
in particular about the mechanical instability of Bohr's orbits -
some results were supported by calculations of Ludwig Föppl (1887-1976)
and mentioned on Sommerfeld's book \cite{Sommerfeld5}. In Nicholson's
model, there were no quantum jumps and the spectral frequencies were
vibration frequencies of the electrons in their positions. Yet this
model gained some notoriety until World War I, thereafter it was set
aside in favor of the Bohr-Sommerfeld model \cite{Kragh2}.

\subsubsection{The photoelectric phenomenon}

Here it is important to note, at first, that at that time there were
several theories to explain the photoelectric effect\cite{Jammer,Stuewer,Wheaton}.
We can cite here the theories of Arthur Erich Haas (1884-1941) and
Joseph John Thomson (1856-1940), before our approach to the Sommerfeld
one.

Haas was the author on 1910 of a model of the hydrogen atom which
allowed him to deduce, for the first time (and three years before
Bohr's deduction\cite{Bohr1,Bohr2,Bohr3}), the Rydberg constant in
terms of the charge and mass of the electron and the Planck constant,
but with a wrong numerical factor \cite{Jammer}. However, in his
theory for the photoelectric effect, Haas used the more known Thomson's
model (where electrons were immersed on a larger positive sphere)
\cite{Jammer} and supposed that, the electron inside the atomic positive
sphere would oscillate when a wave incided over the atom. For an electron
on the boundary of the atomic sphere, its energy would be $h\nu$
and, if the energy of the incident wave was larger than that amount,
the electron would be ejected and the $h\nu$ would be simultaneously
abstracted from the incident wave \cite{Stuewer}. The Haas theory
was supported by Hendrik Antoon Lorentz (1853-1928), who presented
it on a lecture on Göttingen in October 1910 \cite{Stuewer,Wheaton}
and used his own theory of the electron \cite{Lorentz} to analyze
how classical waves might deliver quantum units of energy to bound
electrons of the surface.

Thomson, by this time, propose two distinct theories, associated with
two distinct atomic models, different from his former model too. In
both theories, the emission of the electrons would be by \emph{resonance}
between the frequency of the incident radiation and the characteristic
frequency (associated to revolutions or oscillations) of the electron
in the atom and, in the second one, he reached the relation

\begin{equation}
T=h\nu
\end{equation}
where $T$ is the kinetic energy.

Some other names related to the photoelectric effect were Philipp
Lenard (1862-1947), Carl Ramsauer (1879\textendash 1955), Erich Marx
(1879-1955) and Alexandr Stoletov (1839-1896). It is interesting to
mention, however, the theories of Peter Franken (1928-1999) \cite{Stuewer}
and Owen Richardson (1879-1959)\cite{Wheaton}, both them containing
without further no assumptions about the nature of light. Richardson's
theory was developed on 1914 and had a thermodynamic appeal, treating
only with macroscopic quantities, by analogy between the effect and
the evaporation of molecules from a liquid \cite{Wheaton}. Franken's
theory, curiously, is a contemporary one and uses time-dependent perturbation
theory with Schrödinger's equation. It assumes that the atom is in
its ground state and is subjected to a perturbation represented by
a classical electromagnetic wave of well defined frequency, with this
wave having enough energy to induce over the electron the transition
to continuum. Franken obtained the same expression of Einstein's theory
(including the work function), and Planck's constant $\hbar$ is introduced
by Schrödinger's equation \cite{Stuewer}.

Let us return to the Sommerfeld model to photoelectric effect \cite{Debye1,Sommerfeld6}.
It was innovative because he introduced $h$ from the beginning with
the rule of Eq. (\ref{Sommerfeld}) \cite{Jammer,Stuewer,Wheaton},
contrary to the models of Haas and Thomson, who introduces $h$ at
the end of the calculations. The incident radiation made the electron
oscillates and, in resonance, after a certain time $\tau$, the electron
will be ejected of the atom. This is very similar to the explanation
of Ishiwara in the section C of his paper. It is important to note
that none of the mentioned theories considers the \emph{work function}
on the law for photoelectric effect. For Sommerfeld, this function
was not related to the effect itself, but associated to the path of
the photoelectron from the ejection from atom to the surface of the
metal under analysis. At last, for the Sommerfeld theory, a plot between
photoelectron's energy and the atomic characteristics frequencies
will have, to each atomic frequency, a maximum (Sommerfeld shows some
graphics on his paper with Debye\cite{Debye1}), contrary to Einstein's
theory, where the frequency of the photoelectrons independs of any
atomic frequency\cite{Stuewer}. A deep analysis of Sommerfeld theory
can be found on the chapter 7 of Wheaton's book\cite{Wheaton}.

Then, even more than one decade after Einstein's paper about photoelectric
effect was published in \emph{Annalen der Physik }\cite{Einstein4},
his theory was not the only one, and not widely accepted yet (we may
disregard the little known theories of Franken and Richardson here).
As a matter of fact, this is also due to the lack of precise experimental
data \cite{Jammer,Stuewer,Wheaton}, but the problem was solved by
the verification of Einstein's law (a non-resonance phenomenon with
linear energy-frequency relation for the emitted electrons) by Millikan,
who was working in it at the time of Ishiwara's paper \cite{Milikan1,Milikan2,Milikan3,Milikan4}.
The Nobel Prize for Einstein \cite{Einstein2} \textquotedbl{}for
his services to Theoretical Physics, and especially for his \emph{discovery
of the law of the photoelectric effect}\textquotedbl{} on 1921 - while
he was in Japan - and, for Millikan \cite{Millikan5} \textquotedbl{}for
his work on the elementary charge of electricity and on \emph{the
photoelectric effect}\textquotedbl{}, on 1923. After them, the questions
of Ishiwara on the end of his paper were naturally answered. It is
curious that Millikan refers to the numerical value of the elementary
charge $e$ as \textquotedbl{}the author's value\textquotedbl{}\cite{Milikan1}
or \textquotedbl{}my value\textquotedbl{} \cite{Milikan3}. On \cite{Milikan3},
we have a complete description of the experimental apparatus of Millikan,
a portrait of his ingenuity, and the original graphics of results.

Ishiwara used the Sommerfeld model from 1911 to show the \textquotedbl{}special
power\textquotedbl{} of its formula (\ref{eq1}), considering the
case where the frequency of the incident wave is the same as the natural
frequency of the electron and neglecting the damping. Until electron's
ejection, its movement is described by the forced-damping oscillator;
after its ejection, Ishiwara applies his equation (\ref{eq1}) where
he uses $j=1$, preventing the influences of his incorrect condition.
He obtains the \textquotedbl{}accumulation time\textquotedbl{} in
accordance with the value calculated from the Sommerfeld theory. At
last, Ishiwara obtains the linear relation between the energy and
the frequency of the electron, in accordance with Einstein's law.
However, the potential energy is present on the final expression,
disagreeing with Sommerfeld's theory. Ishiwara purposes the question
\textquotedbl{}Where does the potential energy $U$ remain?\textquotedbl{}

\section{Conclusion}

The work analyzed here is one hundred years old and it take part at
the apex of attempts to understand and explain the spectroscopy data
\cite{Bucher,Castro,Duncan,Eckert,Haar,Jammer,Mehra2,Mehra}. Moreover,
it reflects pioneering efforts to understand the relations between
quantum and classical worlds in a structure where the question of
classicality of nature was more evident, because the set of OQT rules
itself, included explicitly the use of classical mechanics from the
beginning.

The question of the interpretations of quantum theory and the division
of quantum and classical worlds is still subject to discussions \cite{Freire,Jammer2,Laloe,Nikolic,Omnes,Pessoa1,Pessoa2,Neto,Zurek},
but the focus changed with the emergence of concepts \cite{Jammer,Mehra2},
with which any modern physicist is familiar, and were consolidated
10 years after Ishiwara's work - uncertainty relations (Heisenberg
\cite{Zurek}, 1927) , wave particle duality (from Bohr culminating
with the proposal of Louis de Broglie (1892-1987) on 1924\cite{Broglie}),
wave equations (Schrödinger, 1926 \cite{Schrodinger} and Dirac 1928
\cite{Dirac1}) and wave function with its probabilistic interpretations
(Born \cite{Ludwig,Zurek} and Dirac \cite{Dirac2}, both on 1926).
Today, contrary to OQT, the classical physics enters implicitly in
the optical-mechanic analogy (based on the Hamilton-Jacobi theory
\cite{Fetter}) of Schrödinger in the deduction of his equation \cite{Joas,Koberle,Wessels}
and explicitly just in some specific potentials inspired on classical
expressions (e.g., the harmonic and Coulombian ones \cite{Eisberg})
needed to solve the equation.

We hope that this work made clear how rich the first 30 years of QM
were, raising questions rarely seen on textbooks - thermodynamics
with chemical constants, the atomic models of Nagaoka and Nicholson,
the way from Planck to Sommerfeld on phase-space analysis and the
alternative theories for photoelectric effect - and stimulates new
studies about the career of Jun Ishiwara, a physicist who is rarely
remembered but, as we saw on section II, had a wide range of skills.
Indeed, besides his contribution to disseminate the theory of relativity
in the Eastern world, Ishiwara lost the primacy to the proposal of
phase-space quantization rule by just two months (his work was published
on May 1915, and Wilson's work in March of the same year). Though
his paper (and also Wilson's paper) exerted no significant influence
on the further development of the quantum theory of atoms \cite{Kragh2},
Ishiwara rivaled in some aspects with names of stature of Poincaré,
Planck, Bohr, and Sommerfeld, although of course his work lacked the
depth of that of cited ones. 
\begin{acknowledgments}
The authors wish to thank the editors of EPJH and the referees for
helped us to improve our paper.

C. A. Brasil wishes to thank Thaís V. Trevisan (Instituto de Física
'Gleb Wataghin' IFGW / Universidade Estadual de Campinas UNICAMP and
School of Physics and Astronomy / University of Minnesota) and Reginaldo
de J. Napolitano (IFSC/USP) for reading and criticizing the preliminary
versions of this paper, Miled H. Y. Moussa (IFSC/USP) for his hospitality
along 2016 and to Programa Nacional de Pós-Doutorado (PNPD) from Coordenação
de Aperfeiçoamento de Pessoal de Ensino Superior (CAPES).

K. Pelogia would like to thank to Carlos Brasil for this oportunity,
to Dr. Stephan Humeniuk (Universität Stutttgart) for his corrections
and suggestions, to Dr. Klaus Hentschel (Universität Stutttgart) for
his time to help us and to Christian Wilde, Cristal da Rocha (USP)
, James Payne, Sheila Clark: thank you for read and re read this paper
always looking for mistakes and helping us to improve our paper. 
\end{acknowledgments}


\begin{thebibliography}{Bhattacharyya 2006}
\bibitem[Ishiwara 2017]{Ishiwara3}Jun Ishiwara, \textquotedbl{}The
universal meaning of the quantum of action\textquotedbl{}, to be completed
later with data from EPJH.
\end{thebibliography}
\end{document}